# Elemental topological ferroelectrics and polar metals of few-layer materials

Hu Zhang*, Lulu Zhao, RuiFeng Zhang, Chendong Jin, Ruqian Lian, Peng-Lai Gong, RuiNing Wang, JiangLong Wang, and Xing-Qiang Shi

*Key Laboratory of Optic-Electronic Information and Materials of Hebei Province, Research Center for Computational Physics of Hebei Province, Hebei Key Laboratory of High-precision Computation and Application of Quantum Field Theory, College of Physics Science and Technology, Hebei University, Baoding 071002, P. R. China*

* E-mails: zhanghu@hbu.edu.cn

Ferroelectricity can exist in elemental phases as a result of charge transfers between atoms occupying inequivalent Wyckoff positions. We investigate the emergence of ferroelectricity in two-dimensional elemental materials with buckled honeycomb lattices. Various multi-bilayer structures hosting ferroelectricity are designed by stacking-engineering. Ferroelectric materials candidates formed by group IV and V elements are predicted theoretically. Ultrathin Bi films show layer-stacking-dependent physical properties of ferroelectricity, topology, and metallicity. The two-bilayer Bi film with a polar stacking sequence is found to be an elemental topological ferroelectric material. Three and four bilayers Bi films with polar structures are ferroelectric-like elemental polar metals with topological nontrivial edge states. For Ge and Sn, trivial elemental polar metals are predicted. Our work reveals the possibility of design two-dimensional elemental topological ferroelectrics and polar metals by stacking-engineering.

## I. INTRODUCTION

Elemental ferroelectric materials contain only one element and thus may have very simple crystal structures without inversion symmetry. In a work published in 2018, we have proposed a key rule for the design of elemental ferroelectrics, i.e., atoms should occupy at least *two inequivalent* Wyckoff positions in a crystal with a polar space group for an elemental ferroelectric [1]. The crystal has a nonpolar space group if atoms occupy only one equivalent Wyckoff position. There exist charge transfers for atoms of the same element occupied inequivalent Wyckoff positions in elemental solids, in which ionicity of bonding appears. Inequivalent Wyckoff positions induce inequivalent electronegativity for atoms of the same element. Therefore, inequivalent Wyckoff positions make atoms of the same element have different physical and chemical properties and thus play the similar rule of cation and anion in ferroelectrics with multi-elements such as diatomic ferroelectrics GeTe (*R*3*m*), PbS (*P*6$_3$*mc*) [2], and perovskite oxides BaTiO$_3$ [3]. Such design principle is suitable for both elemental ferroelectrics and elemental polar metals (ferroelectric-like materials with metallicity). Elemental polar metals predicted in Ref. [1] were three-dimensional (3D) crystals with group-V elements (P, As, Sb, and Bi) occupied two inequivalent Wyckoff positions in a distorted α-La-like structure with a polar space group *P*6$_3$*mc*. Ionic characters were found in these materials due to the existence of charge transfers. Ionicity of bonding in elemental solids and elemental ferroelectrics are highly connected. In addition, ionicity of bonding in elemental solids was investigated in Ref. [4] with the help of four-class categorization based on the occupation of Wyckoff positions, which is useful to design elemental ferroelectrics and elemental polar metals. Charge transfers between C atoms occupying inequivalent Wyckoff positions in each honeycomb layer of *graphite* were also discovered. We also recognized that charge transfers in elemental solids may be comparable with those in oxides.

In a recent experimental work [5], ferroelectricity was found in a two-dimensional (2D) elemental material Bi in the black phosphorous like structure with the in-plane electric polarization based on a previous theoretically work [6]. The head-to-head and tail-to-tail domain walls were observed. Ferroelectricity was also found recently in mixed-stacking few-layer graphene with out-of-plane polarization [7-9]. 2D ferroelectrics with out-of-plane polarization for bilayer phosphorene, arsenene, and antimonene were predicted [10]. Up to now, various other elemental ferroelectrics were also proposed theoretically [5]. For 2D elemental ferroelectrics, atoms should occupy at least two inequivalent Wyckoff positions in a crystal with a polar *layer* group.

Stacking-engineering has been successfully used to control physical properties of 2D materials in experiments. For trilayer graphene, there are three stacking sequences, rhombohedral (ABC), Bernal (ABA), and simple hexagonal (AAA). All these three different stackings with distinctive electronic structures have been observed experimentally [11]. Furthermore, two methods including Joule heating and laser pulse excitation have been presented to turn the ABC stacking into the ABA stacking controllably and locally in an experimental work in 2018 [12]. The other example is the sliding ferroelectricity [8,13,14], in which vertical polarization induced in their bilayers and multilayers as a result of certain stacking can be switched upon interlayer sliding driven by a vertical electric field. Experimental works have confirmed this mechanism in 2D materials including BN, $MoS_2$, and $MoTe_2$ [15-17]. Due to the advantage of controllability, it is attractive for seeking 2D elemental ferroelectric materials having multifunctional physical properties such as topological and conductive properties by stacking-engineering. We also know that, in analogy with graphene, a single bilayer (BL) Si, Ge, Sn and Bi in the buckled honeycomb lattice were found to be topological insulators in previous theoretical and experimental works [18-23], which motivates our present study.

In this work, we first study the emergence of ferroelectricity in 2D elemental materials through stacking-engineered buckled honeycomb lattices. Various stackings allow atoms occupy multiple inequivalent Wyckoff positions in a polar layer group, which is required for elemental ferroelectrics as discussed above. As a result, the out-of-plane polarization appears in these systems. Then we investigate physical properties of such structures formed by group-IV and group-V elements. Layer-stacking-dependent ferroelectricity, topology, and metallicity are discovered for Bi. 2D elemental polar metals are obtained for Ge and Sn.

## II. METHODOLOGY

First-principles calculations were performed based on density functional theory (DFT) [24] with the Perdew–Burke–Ernzerhoff (PBE) functional in generalized gradient approximation (GGA) [25] in the Vienna Ab Initio Simulation Package (VASP) [26-28]. Electronic structures were checked with the Heyd-Scuseria-Ernzerhof (HSE) hybrid functional [29]. We used a 12×12×1 Monkhorst-Pack grid [30] in PBE calculations and a 6×6×1 Monkhorst-Pack grid in HSE calculations. An energy cutoff of 500 eV was used. All crystal structures were relaxed until the Hellmann-Feynman forces are less than 1 meV/Å. Band structures were calculated with the consideration of spin-orbit coupling (SOC) effects. Tight-binding models were constructed with the maximally localized Wannier functions using Wannier90 package [31,32]. Topological properties were calculated with WannierTools [33].

## III. RESULTS AND DISCUSSION

### A. General characters for 2D elemental ferroelectrics

Firstly, we discuss the general method to design 2D elemental ferroelectrics. Two issues need to be solved to obtain 2D elemental ferroelectrics. The first one is to determine that 2D elemental ferroelectrics should have what crystal structures. Obviously, to be ferroelectrics, 2D elemental materials must have polar layer group only in which electric polarization can exist. The polar layer groups containing six-fold rotation include $p6mm$ (No. 77), $p622$ (No. 76), and $p6$ (No. 73). The polar layer groups containing four-fold rotation include $p4bm$ (No. 56), $p4mm$ (No. 55), $p42_12$ (No. 54), $p422$ (No. 53), and $p4$ (No. 49). The three-fold rotation appears in $p31m$ (No. 70), $p3m1$ (No. 69), $p321$ (No. 68), $p312$ (No. 67), and $p3$ (No. 65). Other polar layer groups have low symmetry. In addition, like 3D elemental ferroelectrics, inequivalent Wyckoff positions must be occupied since this will make atoms of the same element play the similar rule of cation and anion in ferroelectrics with multi-elements. For example, atoms in elemental ferroelectrics with the polar layer group *p6mm* may occupy the following Wyckoff positions with high symmetry: 1a (0,0,$z$) with site symmetry $C_{6v}$, 2b (1/3,2/3,$z$) (2/3,1/3,$z$) with site symmetry $C_{3v}$, and 3c (1/2,0,z) (0,1/2,$z$) (1/2,1/2,$z$) with site symmetry $C_{2v}$. The requirement of occupying inequivalent Wyckoff positions is easily satisfied in 2D elemental materials with multilayer structures. Hence stacking-engineering is an ideal method to obtain 2D elemental ferroelectrics.

According to the definition of ferroelectrics, the other issue is to find out the mechanism of the switching of electric polarization in 2D elemental ferroelectrics under an applied electric field. If a 2D elemental ferroelectric has the monolayer structure, the usual route is relative motions of some atoms in a particular direction. 2D elemental material Bi in the black phosphorous like structure is such example [5]. For 2D elemental ferroelectrics with multilayer structures, relative motions of each layer are also possible means for the switching of electric polarization. This is the mechanism of switching of ferroelectricity in multilayer graphene [8]. In the next section, we apply these results to 2D elemental materials with buckled honeycomb lattices.

### B. Ferroelectricity in stacking-engineered buckled honeycomb lattices

We now consider few-layer elemental materials with various possible stacking sequences of the buckled honeycomb lattice occupied with atoms of the same element in which ferroelectricity may emerge. The side and top views of a single BL buckled honeycomb lattice are shown in Fig. 1(a). This is a centrosymmetric structure with the layer group *p*-3*m*1 (No. 72). For the 2 BLs film, the most common stacking is shown in Fig. 1(b) that can be denoted as the ABCA stacking. The layer group of this structure is also *p*-3*m*1 with four atoms occupying 2b (0,0,$z_1$) (0,0,−$z_1$) and 2c (1/3,2/3,$z_2$) (2/3,1/3,−$z_2$) Wyckoff positions respectively. In this structure, the top (atoms 1 and 2) and bottom (atoms 3 and 4) buckled honeycomb lattices are shifted related to each other as indicated by blue and red hexagons shown in Fig. 1(b). The topological insulator 2 BLs Bi film studied in Ref. [19] has this structure.

The other one is the ABAC stacking as shown in Fig. 1(c). This is an inversion symmetry breaking polar structure with the layer group $p3m1$ (No. 69) in which four atoms occupying 1a (0,0,$z_1$), 1a (0,0,$z_2$), 1b (1/3,2/3,$z_3$), and 1c (2/3,1/3,$z_4$) Wyckoff positions respectively. The ABAC stacking structure comes from the 3D elemental polar metals in a distorted α-La-like structure ($P6_3mc$) discovered in our previous work [1]. The symmetry is very different from 3D to 2D. From the ABCA stacking to the ABAC stacking, the relative $z$ coordinates of atoms 1 and 2 are alternated, which is the key to break the inversion symmetry. For the top BL, atoms 1 and 2 occupy inequivalent Wyckoff positions 1a and 1c respectively, which indicates the existence of charge transfer between them. The same is true for atoms 3 and 4 at the bottom BL. Obviously, multi-inequivalent Wyckoff positions are occupied in this polar layer group, which satisfies the requirement of the formation of 2D elemental ferroelectrics discussed above. The electric polarization will emerge in materials with this stacking structure. Due to the stacking and relative displacements of atoms along the $z$ direction, the electric polarization is out-of-plane different from the experimentally discovered 2D elemental Bi in the black phosphorous like structure with the in-plane electric polarization [5]. Elemental materials having the ABAC stacking would be ferroelectrics if this electric polarization can be switched experimentally.

There is also another stacking denoted as CABA shown in Fig. 1(d). This CABA stacking can be transformed from the ABAC stacking by an inversion operator. In fact, materials with the ABAC and CABA stackings correspond to two symmetry-related polarization states of ferroelectrics and thus have the same energy but an inversed electric polarization as indicated by green arrows in Fig. 1(c, d). The centrosymmetric ABCA stacking phase shown in Fig. 1(b) behaves like a nonpolar high-symmetry paraelectric state. The ferroelectric structure transition is from the ABCA stacking to the polar ABAC and CABA stackings. Fig. 1(e) gives the schematic diagram of expected stacking transitions between these states, in analogy with experimental realized stacking transitions of multi-layer graphene. Comparing structures of ABAC and CABA stackings, the switching of ferroelectricity can be achieved through relative motions of atoms in the $z$ direction under an applied electric field.

For 3 BLs films, we show three typical structures with CA-BD-AB, ABAC-AB, and ABAC-BA stacking sequences shown in Fig. 2(a). The structure of the CA-BD-AB stacking is centrosymmetric and has the layer group $p$-$3m1$ with six atoms occupying 2b $(0,0,z_1)$ $(0,0,-z_1)$, 2c $(1/3,2/3,z_2)$ $(2/3,1/3,-z_2)$, and 2c $(1/3,2/3,z_3)$ $(2/3,1/3,-z_3)$ Wyckoff positions respectively. The inversion symmetry breaking structure with the ABAC-AB stacking has the layer group $p3m1$ with six atoms occupying 1a $(0,0,z_1)$, 1a $(0,0,z_2)$, 1a $(0,0,z_3)$, 1b $(1/3,2/3,z_4)$, 1b $(1/3,2/3,z_5)$, and 1c $(2/3,1/3,z_6)$ Wyckoff positions respectively. The structure of the ABAC-BA stacking is also a polar phase with the layer group $p3m1$, in which relative $z$ coordinates of atoms in the top layer are changed compared to these in the ABAC-AB stacking. Due to such difference in the top layer, the magnitude of electric polarization in ABAC-AB and ABAC-BA stackings should have different values as indicated by arrows with different length in Fig. 2(a).

For 4 BLs films, three typical structures with BD-ABCA-BD, ABAC-ABAC, and ABAC-ABCA stacking sequences are shown in Fig. 2(b). The structure of the BD-ABCA-BD stacking is centrosymmetric and has the layer group $p$-$3m1$ with eight atoms occupying 2b $(0,0,z_1)$ $(0,0,-z_1)$, 2c $(1/3,2/3,z_2)$ $(2/3,1/3,-z_2)$, 2c $(1/3,2/3,z_3)$ $(2/3,1/3,-z_3)$, and 2c $(1/3,2/3,z_4)$ $(2/3,1/3,-z_4)$ Wyckoff positions respectively. The polar ABAC-ABAC stacking can be viewed as the doubled ABAC stacking in the 2 BLs film shown in Fig. 1(c). For the ABAC-ABCA stacking, the top layer is different from those in the ABAC-ABAC stacking. Hence the magnitude of electric polarization in states with ABAC-ABCA and ABAC-ABAC stackings should have different values as the case in ABAC-AB and ABAC-BA stackings. This reveals the manipulation of polarization by stacking-engineering.

For the purpose of this work, we mainly consider films with above stacking sequences. Thicker films with polar structures in which ferroelectricity emerges may be obtained by combination of the 2, 3, and 4 BLs films discussed above. For example, we can stack the 2 BLs ABAC and 3 BLs ABAC-AB stacking to obtain one 5 BLs film (ABAC-ABAC-AB) with the polar layer group. Thicker films have more stacking possibility to host ferroelectricity but it might be hard to control in experiments. For ferroelectrics obtained by stacking-engineering, there exist unique domain walls that different from bulk ferroelectrics or 2D ferroelectrics with a single layer. In Fig. 2(c), we show the schematic illustration of possible domain walls for 3 BLs films. It would be interesting to study physics of such types of domain walls experimentally.

### C. Topological ferroelectrics and polar metals

We first consider Bi films with above discussed stacking sequences. Previous studies have found that centrosymmetric ultrathin Bi films with buckled honeycomb lattices are topological insulators characterized by a nontrivial $Z_2$ number independent of the thickness [19]. These centrosymmetric structures come from the single crystal bismuth with the space group *R*-3*m*. The centrosymmetric films with 2–4 BLs correspond to the 2 BLs ABCA, 3 BLs CA-BD-AB, and 4 BLs BD-ABCA-BD stacking sequences shown in Fig. 1(b) and Fig. 2(a, b). These structures and 1 BL of Bi are all topological insulators according to studies in Ref. [19]. We calculate the electronic band structures of the Bi films with a ABCA stacking based on the PBE functional considering SOC. Our results shown in Fig. 3(a) are identical to that in Ref. [19]. Bi-2BL-ABCA has a narrow indirect band gap of 0.04 eV. A single $Z_2$ topological number ($v$) is used to describe a two-dimensional topological insulator. To calculate the edge states and the $Z_2$ topological number, we obtain a tight-binding Hamiltonian based on maximally localized Wannier functions. The calculated $Z_2$ number for Bi-2BL-ABCA is $v = 1$, which indicates a topologically nontrivial phase. These results are consistent with those in Ref. [19] indicating the reliability of our calculation methods. The zigzag-edge states of Bi-2BL-ABCA are shown in Fig. 3(b). There are topological nontrivial edge states connecting the valence and conduction bands indicated by red arrows. Above the topological states, trivial edge states are also appeared. The trivial metallic edge states are not robust and can be easily removed by surface impurities or defects in principle [19]. Different from the trivial edge states, topological nontrivial edge states are protected by time-reversal symmetry and thus are robust against local perturbations.

Now we consider the inversion symmetry breaking polar phase of 2 BLs Bi with the ABAC stacking. The PBE electronic band structures calculated with SOC are shown in Fig. 3(c). Bi-2BL-ABAC is a metal due to the overlap of bands at the Fermi level giving a negative band gap of 0.08 eV. As Bi is a heavy element, materials containing Bi always have large SOC effects. Different from band structures of Bi-2BL-ABCA shown in Fig. 3(b), there is giant spin-splitting as a result of large SOC effects and inversion symmetry breaking in the ABAC stacking. The calculate $Z_2$ number for Bi-2BL-ABAC is $v=1$. Thus, the polar phase is still topologically nontrivial like the centrosymmetric phase. The edge states of Bi-2BL-ABAC are shown in Fig. 3(d). We can find both topological nontrivial edge states and trivial edge states similar to Bi-2BL-ABCA.

It is well known that the PBE functional usually underestimates the band gap. To check our results, we also use the HSE06 functional that gives more accurate band gaps in general. The HSE06 electronic band structures of Bi-2BL-ABAC are shown in Fig. 4(a). In this case, Bi-2BL-ABAC is an insulator with a narrow indirect band gap of 0.07 eV. The calculate $Z_2$ number for Bi-2BL-ABAC is still $v=1$. As can be found in Fig. 4(b), topological nontrivial edge states survive. The HSE06 functional does not change the result of topological properties. Hence, Bi-2BL-ABAC is a topological insulator at the HSE06 level.

As discussed above, four Bi atoms in Bi-2BL-ABAC occupy 1a $(0,0,z_1)$, 1a $(0,0,z_2)$, 1b $(1/3,2/3,z_3)$, and 1c $(2/3,1/3,z_4)$ Wyckoff positions respectively. Based on the Bader analysis [34], Bader charges for Bi1~Bi4 (see Fig. 1 for the atomic number 1~4) in Bi-2BL-ABAC are 4.991, 5.012, 4.987, and 5.008 electrons respectively. Clearly, there are small charge transfers between Bi atoms due to occupations of inequivalent Wyckoff positions. As a result, ionicity of bonding in Bi-2BL-ABAC with the layer group $p3m1$ induces ferroelectricity [1]. The energy difference between polar Bi-2BL-ABAC and nonpolar Bi-2BL-ABCA is 0.04 eV. This value can be used to estimate the energy barrier of realistic switching of the electric polarization. The energy barrier is sufficiently small for switching. If this is confirmed in experiment, Bi-2BL-ABAC would be an elemental topological ferroelectric material.

Next, we consider thicker Bi films with inversion symmetry breaking stacking sequences shown in Fig. 2. The PBE electronic band structures of Bi-3BL-ABAC-AB calculated with SOC are shown in Fig. 5(a). One can find giant spin-splitting due to large SOC effects and inversion symmetry breaking. The large overlap of bands at the Fermi level gives a band gap −0.29 eV. Hence Bi-3BL-ABAC-AB is a polar metal. The calculate $Z_2$ number is $\nu = 1$ indicating a topological character. Topological nontrivial and trivial edge states appear in Fig. 5(b). For Bi-3BL-ABAC-BA, PBE band structures are plotted in Fig. 5(c). The band gap is −0.14 eV that is reduced by half compared to that of the ABAC-AB stacking. This reveals stacking-dependent metallic properties. From the PBE band structures of Bi-4BL-ABAC-ABAC shown in Fig. 5(d), we know that it is a polar metal with a band gap of −0.42 eV. The band gap of Bi-4BL-ABAC-ABCA is −0.30 eV. The band overlap of Bi-4BL phases near the Fermi level is modified by the intrinsic electric polarization as a result of different polar stacking sequences. Our calculations also indicate that Bi films with the ABAC-BA, ABAC-ABAC, and ABAC-ABCA stacking sequences all have topological nontrivial edge states. Calculations with HSE06 do not change the metallic and topological nature of Bi-3BL and Bi-4BL films. Therefore, we conclude that polar states of Bi-3BL and Bi-4BL films are all elemental polar metals having topological nontrivial edge states.

With the help of stacking-engineering, we obtain elemental topological ferroelectrics and polar metals in ultrathin Bi films. Results are collected in Fig. 5(e). These states show layer-stacking-dependent physical properties. Up to now, experimental grown ultrathin Bi films are centrosymmetric. It is desired to make efforts to grow inversion symmetry breaking polar states of Bi discussed in this work. Similar to multilayer graphene and sliding ferroelectrics, turning structures of Bi films among different stacking sequences might be achieved experimentally. In addition, the electric polarization in some 2D polar metals such as $WTe_2$ has been switched in experiment [35]. Hence it is interesting to check the switchability of electric polarization in elemental polar metals predicted here.

Finally, we investigate 2BL-ABAC films with polar symmetry formed with other elements. The PBE electronic band structures are shown in Fig. 6. There is an energy band gap of 0.27 eV for Sb-2BL-ABAC. While Sn, Ge, and Si with the ABAC stacking are all metals. As noted above, Sn, Ge, and Si 1BL films are topological insulators. However, the calculated $Z_2$ number is $v = 0$ for Sn-2BL-ABAC, Ge-2BL-ABAC, and Si-2BL-ABAC and thus there are no topological nontrivial edge states. For HSE06 calculations, Sn-2BL-ABAC and Ge-2BL-ABAC are still trivial polar metals, and Si-2BL-ABAC has a small gap of 0.06 eV without nontrivial edge states. Comparing to their single bilayer phases, polar stacking sequences changes topological properties. Unlike Bi, the stacking-engineering has different influence on topological properties of group IV elements. Bader charges for Si1~Si4 in Si-2BL-ABAC are 4.034, 3.962, 3.831, and 4.171 electrons respectively. Charge transfers between Si atoms are larger than those for Bi atoms in Bi-2BL-ABAC.

## IV. CONCLUSIONS

In conclusion, we construct 2D elemental phases with various possible stacking sequences of the buckled honeycomb lattice in which ferroelectricity may emerge. In ferroelectric states, atoms occupy at least two inequivalent Wyckoff positions in crystals with a polar layer group. Through stacking-engineering, the elemental topological ferroelectric is designed for two-bilayer Bi. Elemental polar metals with nontrivial edge states are discovered for ultrathin Bi films with different stackings. For group IV elements Ge and Sn, trivial elemental polar metals are found. Stacking-engineering is a powerful method to design elemental materials with the coupled physical properties of ferroelectricity, topology, and metallicity.

## ACKNOWLEDGMENTS

This work was supported by the Advanced Talents Incubation Program of the Hebei University (Grants No. 521000981423, No. 521000981394, No. 521000981395, and No. 521000981390), the Natural Science Foundation of Hebei Province of China (Grants No. A2021201001 and No. A2021201008), the National Natural Science Foundation of China (Grants No. 12104124 and No. 12274111), the Central Guidance on Local Science and Technology Development Fund Project of Hebei Province (236Z0601G), Scientific Research and Innovation Team of Hebei University (No. IT2023B03), and the high-performance computing center of Hebei University.

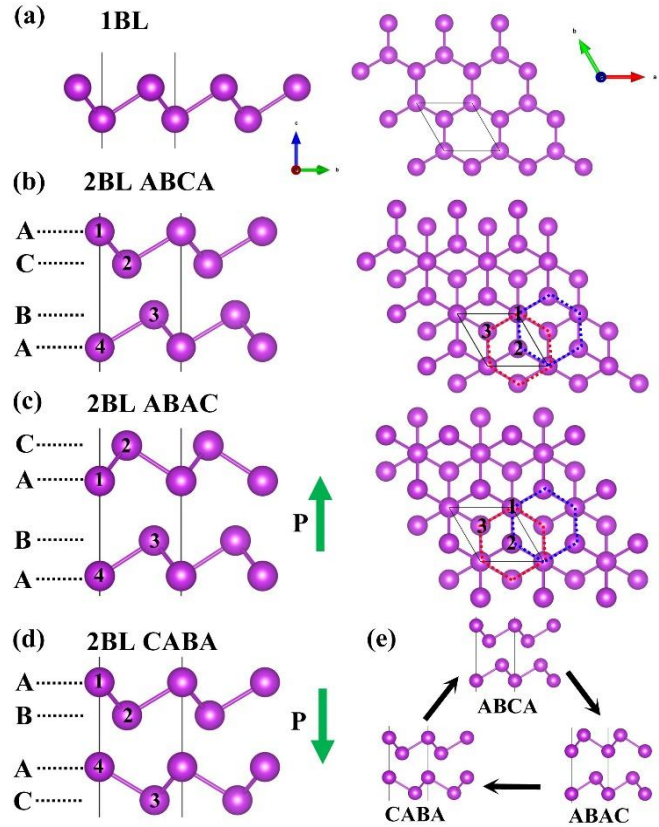

FIG. 1. The side and top views of (a) a single bilayer (BL) buckled honeycomb lattice and 2BL with the (b) ABCA, (c) ABAC, and (d) CABA stackings. (e) Schematic illustration of tuning among different stackings. For each stacking, we use a capital letter to denote each layer from the bottom to top. The CABA stacking can be transformed from the ABAC stacking by an inversion operator. The green arrows indicate the electronic polarization. The honeycomb formed by top atoms (1 and 2) and bottom atoms (3 and 4) are denoted by blue and red hexagons respectively.

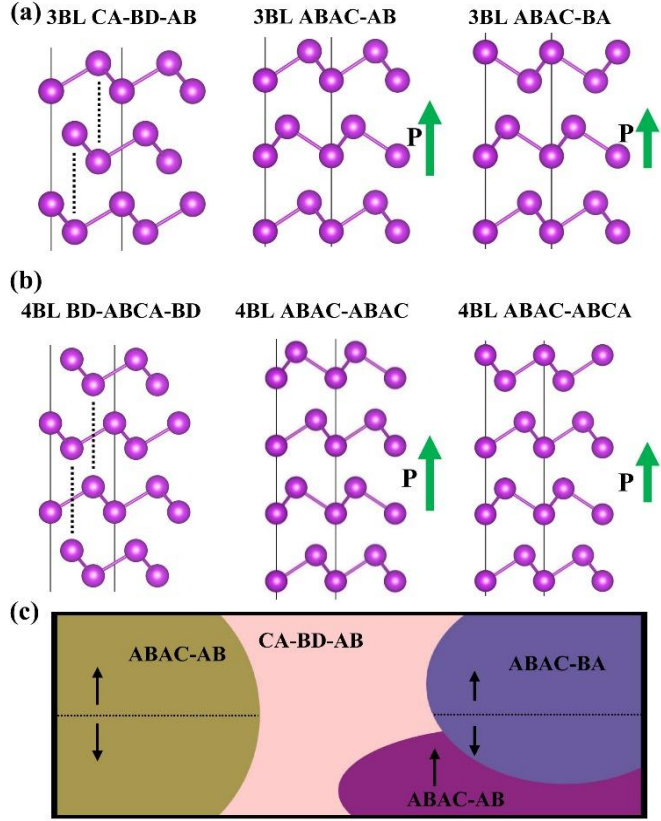

FIG. 2. The side view of (a) 3BL and (b) 4BL buckled honeycomb lattice with various stackings. (c) Schematic illustration of possible domain walls for materials of 3BL buckled honeycomb lattice with different stacking structures denoted as areas with different colors. For each stacking, we use a capital letter to denote each layer from the bottom to top. The green and black arrows indicate the electronic polarization.

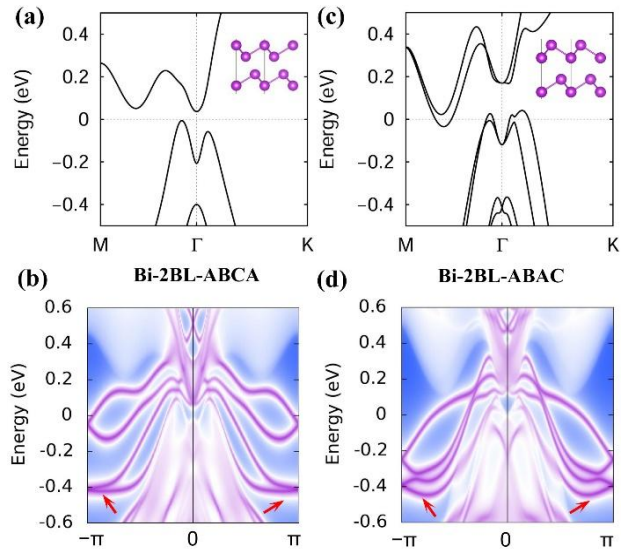

FIG. 3. PBE band structures of (a) Bi-2BL-ABCA and (c) Bi-2BL-ABAC calculated with SOC. Zigzag-edge spectral functions of (b) Bi-2BL-ABCA and (d) Bi-2BL-ABAC. Topological nontrivial edge states are indicated with red arrows.

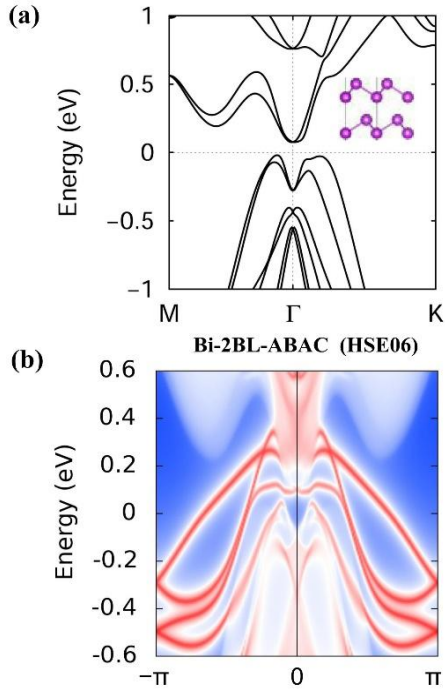

FIG. 4. (a) Band structures and (b) zigzag-edge spectral functions of Bi-2BL-ABAC calculated with the HSE06 functional.

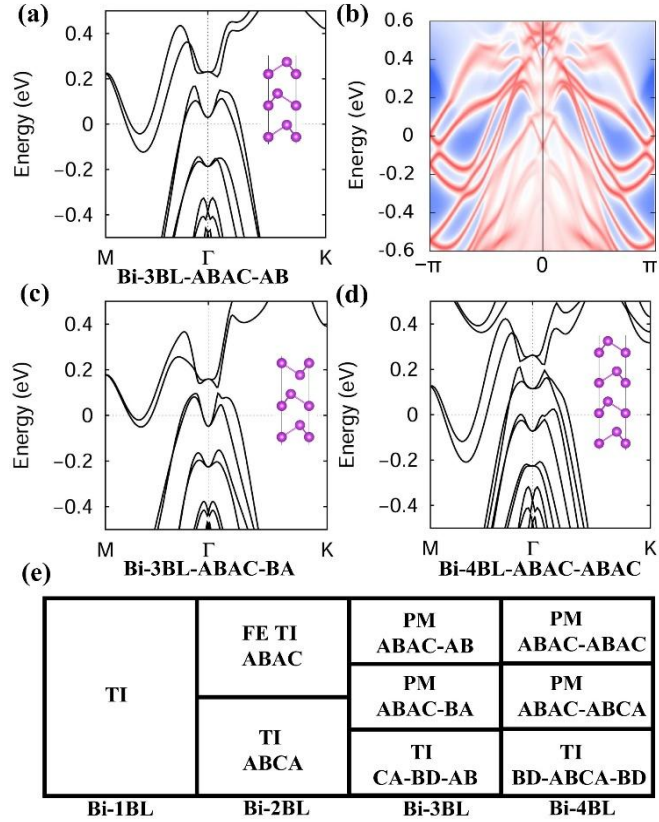

FIG. 5. PBE band structures of (a) Bi-3BL-ABAC-AB, (c) Bi-3BL-ABAC-BA, and (d) Bi-4BL-ABAC-ABAC. (b) Zigzag-edge spectral functions of Bi-3BL-ABAC-AB. (e) Layer-stacking-dependent properties of ultrathin Bi films for the topological insulator (TI),

ferroelectric (FE), and polar metal (PM) states.

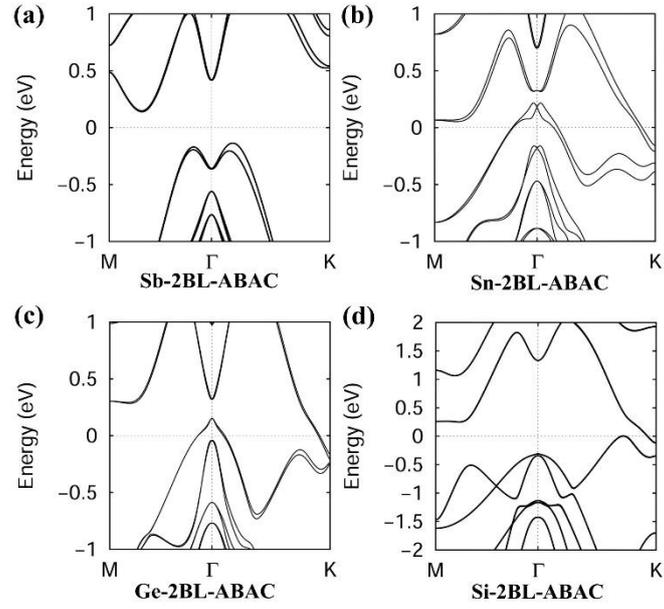

FIG. 6. PBE band structures of (a) Sb-2BL, (b) Sn-2BL, (c) Ge-2BL, and (d) Si-2BL with the ABAC stacking.